\documentclass[final,3p,times,twocolumn]{elsarticle}

\usepackage{sistyle}
\usepackage[version=3]{mhchem}
\usepackage{xspace}
\usepackage{graphicx}
\usepackage{multirow}
\usepackage{setspace}
\usepackage[usenames,dvipsnames]{color}

\usepackage{amssymb}

\newcommand{\fref}[1]{Fig.~\ref{#1}}
\newcommand{\sref}[1]{Sec.~\ref{#1}}

\newcommand\etal{\textit{et al.} }

\newcommand\Ar{\ce{^{31}Ar}\xspace}
\newcommand\Si{\ce{^{28}Si}\xspace}
\renewcommand{\S}{\ce{^{30}S}\xspace}
\newcommand\Cl{\ce{^{31}Cl}\xspace}
\renewcommand\P{\ce{^{29}P}\xspace}
\newcommand\Qop{$Q_{1p}$\xspace}
\newcommand\Qep{$Q_{3p}$\xspace}
\newcommand\Qtp{$Q_{2p}$\xspace}
\newcommand\BGT{$B_{GT}$\xspace}

\newcommand\SIE[3]{\num{#1}(\num{#2})\,\SI{}{#3}}

\journal{Physical Letters B}

\begin{document}

\begin{frontmatter}

\title{Sizeable beta-strength in \Ar($\beta$3p) decay}

\author[A-DK]{G. T. Koldste}
\author[B-F]{B. Blank}
\author[M-E]{M. J. G. Borge}
\author[M-E]{J. A. Briz}
\author[M-E]{M. {Carmona-Gallardo}}
\author[M2-E]{L. M. Fraile} 
\author[A-DK]{H. O. U. Fynbo}
\author[B-F]{J. Giovinazzo}
\author[A-DK]{J. G. Johansen\fnref{cD-D}}
\author[J-F]{A. Jokinen}
\author[G-S]{B. Jonson}
\author[B-F]{T. {Kurturkian-Nieto}}
\author[G-S]{T. Nilsson}
\author[M-E]{A. Perea} 
\author[M-E]{V. Pesudo}
\author[M-E,H-CR]{E. Picado}
\author[A-DK]{K. Riisager}
\author[J-F]{A. Saastamoinen\fnref{cT-USA}}
\author[M-E]{O. Tengblad}
\author[G-F]{J.-C. Thomas}
\author[CERN]{J. {Van de Walle}}

\address[A-DK]{Department of Physics and Astronomy, Aarhus University, DK-8000 Aarhus C, Denmark}
\address[B-F]{Centre d'{\'E}tudes Nucl{\'e}aire de Bordeaux-Gradignan, CNRS/IN2P3 -- Universit{\'e} Bordeaux I, F-33175 Gradignan Cedex, France}
\address[M-E]{Instituto de Estructura de la Materia, CSIC, E-28006 Madrid, Spain}
\address[M2-E]{Grupo de F{\'i}sica Nuclear, Universidad Complutense, CEI Moncloa, E-28040 Madrid, Spain}
\address[J-F]{Department of Physics, University of Jyv{\"a}skyl{\"a}, FIN-40351 Jyv{\"a}skyl{\"a}, Finland}
\address[G-S]{Fundamental Fysik, Chalmers Tekniska H{\"o}gskola, S-41296 G{\"o}teborg, Sweden}
\address[H-CR]{Secci{\'o}n de Radiaciones, Universidad Nacional, Heredia, Costa Rica}
\address[G-F]{GANIL, CEA/DSM-CNRS/IN2P3, F-14076 Caen Cedex 5, France}
\address[CERN]{CERN, CH-1211 Geneva 23, Switzerland}

\fntext[cD-D]{Present address: Institut f{\"u}r Kernphysik, Technische Universit{\"a}t Darmstadt, D-64289 Darmstadt, Germany}
\fntext[cT-USA]{Present address: Cyclotron Institute, Texas A\&M University, College Station, TX 77843-3366, USA}

\begin{abstract}
  We present for the first time precise spectroscopic information on the recently discovered decay mode $\beta$-delayed 3p-emission. The detection of the 3p events gives an increased sensitivity to the high energy part of the Gamow-Teller strength distribution from the decay of $^{31}$Ar revealing that as much as 30\% of the strength resides in the $\beta$3p decay mode. A simplified description of how the main decay modes evolve as the excitation energy increases in \Cl is provided.
\end{abstract}


\end{frontmatter}


\section{Introduction}
\label{intro}
When the dripline is approached the energy window for $\beta$-decays increases so that delayed nucleon emission eventually becomes possible. The most intensely studied decay mode is $\beta$-delayed proton emission, since protons are easy to detect and the dripline lies closer to stability for proton-rich nuclei. For a review of $\beta$-delayed particle emission see Ref. \cite{b2prev,KarstenRev,Borge2013}. \Ar is one of the most studied $\beta$-delayed proton emitters on the dripline. During the past three decades work on \Ar has focussed on the physics of $\beta$-delayed one- and two-proton emission; here we demonstrate how the decay of \Ar can also be used to elucidate the mechanism of $\beta$-delayed three-proton decay.

The $\beta$3p-decaymode was only recently discovered in the decay of \Ar \cite{Ar3p} and it has previously only been observed in two other cases; \ce{^{45}Fe} \cite{Fe3p} and \ce{^{43}Cr} \cite{Cr3p,Audirac2012}. The information on the decaymode from these experiments is limited to the identification of its existence, since they all used Time Projection Chambers in which the energy resolution is low for four body decays.

With a compact setup for particle detection consisting of six double sided Si strip detectors, we have for the first time obtained precise energy and angular information on a sample of $\beta$-delayed 3p-decays. This enables both a quantitatively and qualitatively new understanding of the decaymode including a significantly improved determination of the high energy part of the Gamow-Teller strength distribution in the decay of \Ar. 

When $\beta$2p-emission is allowed, decays to the Isobaric Analogue State (IAS) will dominate for nuclei closer to stability. Most of the early analysis of decay mechanisms focussed on this, for example the theoretical treatment of $\beta$2p-emission by D\'etraz \cite{Detraz}. We now know that this picture is incomplete, since previous \Ar experiments have shown that many levels in the $\beta$-daughter contribute significantly.  Hence, we need to expand conceptually upon the picture used by D\'etraz and allow for more states to be populated in the $\beta$-daughter. Proton emitting levels close to a threshold tend to give distinct lines, whereas levels at higher excitation energy decay into so many channels that a statistical approach may become more appropriate. The same effect makes $\beta$p-spectra significantly smoother at high $Z$, see e.g. Ref. \cite{Borge2013}. We shall here consider a simplified model, with the aim of focussing on the conceptual issues rather than providing a detail description of the decay.

It was realised early \cite{Ghoshal1950,Morrison1953} for nuclei close to stability that one-neutron emission, two-neutron emission etc. decay modes follow each other as dominating decay modes as the excitation energy in a nucleus is increased. It is thus natural to investigate if a similar sequence appears for $\beta$-delayed proton emission. Here the \Ar decay constitutes an ideal case because it has a sizeable Q-value for three-proton emission, while e.g. the neutron channel only opens above the $Q_{EC}$-window of $\SIE{18.38}{10}{MeV}$ \cite{no2}.

In this letter we first describe our experimental data and give an overview of what information the detailed 3p-spectroscopy can provide. We then focus on the Gamow-Teller distribution and the evolution of decay modes as excitation energy increases.

\section{Three-proton detection} 
\label{data}
The experiment was performed at ISOLDE, CERN using a $\SI{60}{keV}$ beam of \Ar. The beam was collected in a $\SI{50}{\mu g / cm^2}$ carbon foil situated in the center of six double sided silicon strip detectors (DSSSD) (the detector thicknesses were $\SI{69}{\mu m}$, $\SI{494}{\mu m}$, and four with thicknesses around $\SI{300}{\mu m}$). Most of the DSSSDs were backed by thick unsegmented Si detectors. Outside the chamber two Miniball germanium cluster detectors were situated for detection of $\gamma$-rays. A detailed description of the setup, calibration and cuts used in the data analysis can be found in Ref. \cite{no2,no1}. $\beta$-particles deposit of the order of 300 eV/$\mu$m in Silicon while the energy of most protons is in the MeV range, and so the discrimination between protons and $\beta$-particles is mainly based on the deposited energy in the detector. Unless otherwise stated we use a low energy cut-off of 800 keV for the protons, but allow the lowest energy proton in multi-proton events to have as little as 500 keV unless it is detected by the thick strip detector, where the $\beta$-particles can deposited somewhat higher energy than that.

\begin{figure}[tbp]%
	\centering
	\includegraphics[width=0.95\columnwidth]{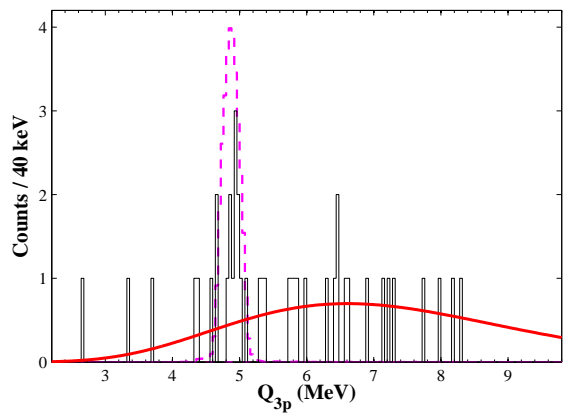}
	\caption{(Colour online) \Qep calculated for multiplicity-three events. The dashed histogram is a simulation of a 3p-decay of the IAS in \Cl to the ground state in \Si. The full curve is the $\chi^2$-distribution from \fref{Q1fig} folded with itself three times and scaled manually to fit the data.}%
	\label{Q3fig}%
\end{figure} 

Ref. \cite{no2} contains a detailed analysis of the Fermi strength of the decay of \Ar and the different decay modes of the IAS. This includes the contribution from the 3p-decay of the IAS, which is clearly identified in the \Qep-spectrum extracted from the multiplicity-three events shown in \fref{Q3fig}. Included is also (purple/grey dashed histogram) a simulation of the decay from the IAS to the ground state (downscaled) to show the energy resolution of the 3p-detection. The full width at half maximum (FWHM) for 3p-decays found from simulations is approximately $\SI{0.3}{MeV}$, mainly due to the reconstruction of the energy of the recoiling \Si nucleus. The resolution for detection of individual protons is $\SI{0.05}{MeV}$ FWHM. It is clearly seen in \fref{Q3fig} that only about half of the events stem from decays of the IAS. The rest is interpreted as coming from levels with excitation energies mainly above the IAS. This implies that there is a contribution to the Gamow-Teller strength which needs to be extracted by a detailed analysis of the $\beta$3p-events. This enables for the first time the inclusion of this decay mode in the determination of the Gamow-Teller distribution, which is done in \sref{GT}. 

The precise energy and angular resolutions also make it possible to study the decay mechanism of the 3p-emission. A detailed discussion of this is also found in Ref. \cite{no2}. It is seen that all the decays can be explained as sequential, but that a simultaneous component cannot be excluded. Decays through a state in \P at $\SIE{3447.6}{4}{keV}$, just $\SI{699}{keV}$ above the proton threshold, is clearly identified for the IAS decays and strong indications of transitions through the next level at $\SIE{4080.5}{3}{keV}$ are also seen. For the levels above the IAS it is not possible to unambiguously determine which states in \P are populated, but if all the decays are sequential it is seen that the decays must go through levels in \P that lie even higher than the two appearing in the IAS decays.

\section{The Gamow-Teller distribution}
\label{GT}
The extraction of the Gamow-Teller strength is much more complicated than the extraction of the Fermi strength, since one has to include all decay modes of the many levels populated in \Cl. It has been attempted previously in an experiment similar to this by Fynbo \etal \cite{Fynbo2p}. In the present experiment the energy and angular resolutions are significantly improved and we can now for the first time include the 3p-component. Disregarding the Fermi contribution one gets for the Gamow-Teller strength:
\begin{align*}
	B_{GT} = \left(\tfrac{g_A}{g_V}\right)^{-2}\frac{C\cdot b}{f\cdot T_{1/2}}, 
\end{align*}
where $C=6144.2(16)$s and $g_A/g_V=-1.2694(28)$ \cite{val}, the $f$-value can be found using Ref. \cite{Wilkinson1974}, the total half-life, $T_{1/2}=\SIE{15.1}{3}{ms}$, determined using these data \cite{no2} and the branching ratio, $b$, can be deduced from the data using the \Qop-, \Qtp- and \Qep-spectra.

To make a precise determination of the branching ratio, it is important not only to have a good energy resolution, but also to precisely know the detection efficiency and the total number of collected \Ar isotopes. Therefore only two of the detectors are used for the \Qtp-spectrum as explained in Ref. \cite{no2}. For the \Qop-spectrum only one of these detectors is used, due to uncertainties in the energy calibration in the other detector for high energies. Since the number of 3p-events is so limited, all detectors are included in the \Qep-spectrum.

\begin{figure}[tbp]%
	\centering
	\includegraphics[width=0.95\columnwidth]{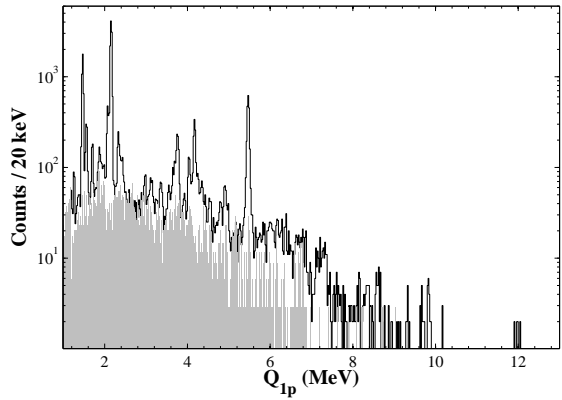}
	\caption{The \Qop-spectrum calculated for all the particles that have hit detector 3 (see Ref. \cite{no2} for the detector placing and numbering). Also shown is the spectrum (grey) calculated for multiple-two events in detector 3 and 6 scaled by the difference in efficiency.}%
	\label{Q1pfig}%
\end{figure} 

The \Qop-spectrum is shown in \fref{Q1pfig}. A lower cut-off of $\SI{1}{MeV}$ is used since the spectrum below is dominated by background. There is a contribution from 2p-decays to the multiplicity-one proton spectrum, which much be subtracted when the \BGT distribution from events leading to bound states in \S is estimated. 
There are also a number of peaks known to stem from decays of the IAS, see Ref. \cite{no2}, which are also removed. Finally some transitions are known not to feed the ground state of \S. The peaks known to go to excited levels in \S below $\SI{5.2}{MeV}$ (listed in Table 2 of Ref. \cite{Fynbo2p}) are shifted upwards with the excitation energy of $\SI{2210.2}{keV}$, $\SI{3404.1}{keV}$, $\SI{3673}{keV}$ and $\SI{5136}{keV}$. Notice that an average energy ($\SI{3673}{keV}$) for the third ($0^+$) and fourth ($1^+$) level is used, since their energies are so close to each other that it is not possible to deduce which one is populated. The branching ratio is then found bin by bin and the \BGT strength deduced. 

For the \Qtp-spectrum a low-energy cut-off of $\SI{500}{keV}$ is used for both protons (the spectrum can be seen in Ref. \cite{no2} Fig. 8). For the extraction of the \BGT strength the three major peaks from the IAS at $\SI{7.6}{MeV}$, $\SI{6.3}{MeV}$ and $\SI{5.7}{MeV}$ are removed by setting the \BGT-value in these bins to be the average of the values in the bins on each side of the peaks. All the decays are assumed to go to the ground state in \P. This we now know not to be true \cite{no2}, but it is not possible to deduce which decays feed which states in \P. In this way we underestimate the \BGT, since if the events go to excited levels, they will come from higher lying levels in \Cl, which due to the $f$-factor will give them a larger \BGT-value.

Finally, the 3p-data used are shown in \fref{Q3fig}. 
Here the decays from the IAS are the events between $\SI{4.3}{MeV}$ and $\SI{5.5}{MeV}$. Thus only events above $\SI{5.5}{MeV}$ are used for the extraction of the \BGT strength. Again all events are assumed to go to the ground state of \Si.

\begin{figure}[tbp]%
	\centering
	\includegraphics[width=0.95\columnwidth]{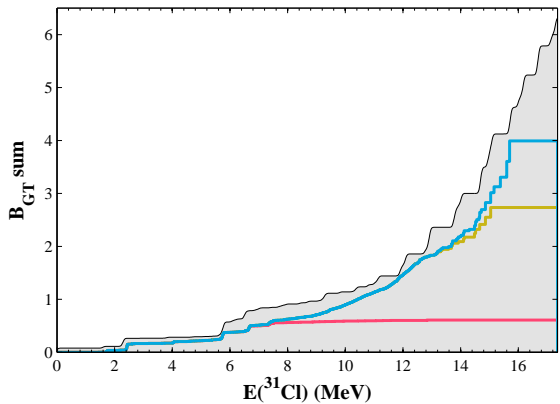}
	\caption{(Colour online) The total summed Gamow-Teller strength from 1p-, 2p- and 3p-events (blue upper curve) as a function of excitation energy in \Cl in the full $Q_{\beta}$-energy window ($\SIE{17.36}{10}{MeV}$). Also shown are the summed \BGT-spectrum from the 1p-events (pink lower curve) and for both the 1p- and 2p-events (yellow middle curve). The shaded area shows the summed Gamow-Teller strength from a shell-model calculation in the full $sd$-shell \cite{Fynbo2p}.}%
	\label{BGTsum}%
\end{figure} 

The total summed $B_{GT}$-spectrum obtained from the three spectra can be seen in \fref{BGTsum} (blue upper curve). The individual contributions from the 1p-events (pink lower curve) and for the sum of the 1p- and 2p-events (yellow middle curve) are also shown. The total observed Gamow-Teller strength in the 1p-spectrum is $0.61(3)(7)$ compared to $0.58(3)$ found in Ref. \cite{Fynbo2p}. The uncertainty is composed of two contributions, where the latter is the $\SI{11}{\%}$ uncertainty on the absolute branching ratio of the main 1p-peak \cite{Axelsson98b} used to extract the number of \Ar ions in the experiment. This uncertainty was not correctly included in Ref. \cite{Fynbo2p}. In the 2p-spectrum the total observed Gamow-Teller strength is $2.1(3)(2)$ compared to $1.86(12)$ in Ref. \cite{Fynbo2p} and in the 3p-spectrum it is $1.26(15)(14)$. As already mentioned, there is a systematic bias for the extracted strength to underestimate the true strength because all transitions are assumed to populate the ground state in the 1p, 2p or 3p daughters unless population of excited states is explicitly measured from coincident gamma rays.  In \fref{BGTsum} the total observed Gamow-Teller strength is compared with a shell-model calculation in the full $sd$-shell model space ($0d_{5/2}$, $1s_{1/2}$ and $0d_{3/2}$) using the Wildenthal universal 11 (USD) effective interaction \cite{Wildenthal1984}, see \cite{Fynbo2p} for further details. The experimental and theoretical curves follow each other closely, indicating that most of the Gamow-Teller strength has been measured up to about $\SI{15.7}{MeV}$. We note that the missing strength above about $\SI{15.5}{MeV}$ in the data corresponds to only 1-2 3p-decay events, so that our results are compatible with the theoretical Gamow-Teller strength in the complete $Q_{\beta}$-window. The one-proton decay mode contributes up to $\SI{12.85}{MeV}$ and the two-proton decay mode starts contributing at $\SI{5.90}{MeV}$, but it can be seen that from around $\SI{7.5}{MeV}$ the contribution from the two-proton decay mode dominates. The two-proton decay mode contributes until $\SI{15.04}{MeV}$ and the three-proton decay mode starts contributing at $\SI{13.16}{MeV}$ and dominate above $\SI{15.04}{MeV}$. From this it is already seen that there is an evolution in the dominating decay modes, i.e. more protons are emitted with increasing excitation energy, as is known to be the case for neutron emission from nuclei close to stability \cite{Ghoshal1950,Morrison1953}. This behaviour is investigated in more detail in the following section.

\section{Evolution of decay modes}
\label{evolution}
The energy distribution of the emitted protons from a given level depends on three factors: The phase-space, which grows proportional to the proton energy, the level density in the daughter nucleus, which increases with excitation energy and the penetrability of the proton, which decreases strongly with decreasing energy~\cite{Hornshoj1972B}. At low proton energy the spectrum is dominated by the penetrability and we expect decay branches to be suppressed if the penetrability is less than $10^{-3}$. Since \Ar has spin $\frac{5}{2}^+$, we only populate $\frac{3}{2}^+$, $\frac{5}{2}^+$ and $\frac{7}{2}^+$ levels in \Cl and given the structure of \Ar we expect the emitted protons to be predominantly from the $sd$-shell, and hence $l=0,2$ to be favoured in the proton emissions. The penetrability of $10^{-3}$ corresponds to proton energies of around $\SI{0.8}{MeV}$ for $s$-wave transitions and around $\SI{1.3}{MeV}$ for $d$-wave transitions for the isotopes involved in the decay and we thus expect decay branches with proton energies below this to be suppressed. For high energies the level density will be so high that the decay can be treated purely statistically \cite{Borge2013}. In Ref. \cite{Detraz} the intensity of protons as a function of energy is calculated and plotted for the decay of the IAS taking into account all three factors. The result is a bell-shaped curve, with a long upper tail. Expanding this to the case here, where multiple levels in \Cl are populated, the energy distribution should be a sum of similar curves, shifted slightly in energy. The result is likely to be a curve of similar shape. This will also be the case for decays of excited levels in both \P and \S, and thus also when one considers the two- and three-proton energy spectra. However, for levels close to the threshold this way of thinking breaks down, and the decay will instead give a number of resolved lines in the spectrum. We now explore whether our data, apart from the resolved lines, can be described by assuming all emitted protons to stem from the same distribution.

\begin{figure}[tbp]%
	\centering
	\includegraphics[width=0.95\columnwidth]{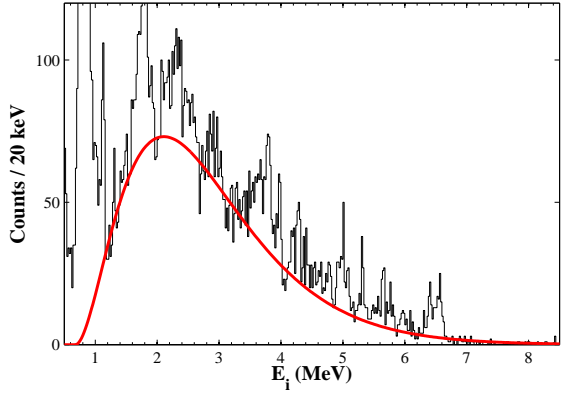}
	\caption{(Colour online) Energies of the individual protons detected in multiplicity-two events. The upper part of the spectrum has been cut off for better visualisation of the structure, when disregarding the resolved lines. The red/grey curve is a $\chi^2$-distribution adjusted to the data manually: $h_{2p}=610\cdot\chi^2(2.8\cdot \tfrac{E}{\SI{}{MeV}}-2.0,6)$.}%
	\label{E2pfig}%
\end{figure}

\begin{figure}[tbp]%
	\centering
	\includegraphics[width=0.95\columnwidth]{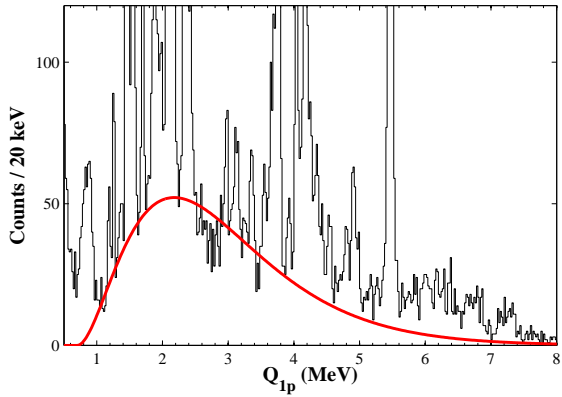}
	\caption{(Colour online) \Qop calculated for all particles detected in detector 3. The upper part of the spectrum has been cut off for better visualisation of the structure, when disregarding the resolved lines. The red/grey curve is the $\chi^2$-distribution from \fref{E2pfig} transformed using the recoil factor and scaled manually to fit the data.}%
	\label{Q1fig}%
\end{figure} 

We first consider the energy spectrum for two-proton events in \fref{E2pfig}. 
The reason for starting with the two-proton spectrum is that the levels fed in \S have a sufficiently large excitation energy so that the level density is high. Therefore this spectrum contains fewer resolved lines than the one-proton spectrum. The major resolved lines around $\SI{0.7}{MeV}$ are most likely due to an overlap in the structure of \Ar and the $\SI{5.2}{MeV}$ state fed in \S. Disregarding the resolved lines in the spectrum the energy distribution does seem to have the same shape as calculated by D\'etraz in Ref. \cite{Detraz}, i.e. a bell-shape with a long upper tail. A well known function with this shape is the $\chi^2$-distribution. By manually adjusting the parameters of this function to the data the following function results
\begin{align*}
	h_{2p} &= 610\cdot\frac{1}{2^{\tfrac{6}{2}}\Gamma\left(\tfrac{6}{2}\right)}x^{\tfrac{6}{2}-1}e^{-\tfrac{x}{2}} \\
	 x&=2.8\cdot \tfrac{E}{\SI{}{MeV}}-2.0,
\end{align*}
which is shown together with the data in \fref{E2pfig}. The distribution has an average energy of $\SI{2.86}{MeV}$.

Multiplying the energy with the recoil factor one gets the typical value of \Qop. This is scaled and plotted in \fref{Q1fig} together with the \Qop-value for all particles which hit detector 3 (the same detector as used in \sref{GT}). Again disregarding the resolved lines, the shape of the data seem to fit the distribution.

\begin{figure}[tbp]%
	\centering
	\includegraphics[width=0.95\columnwidth]{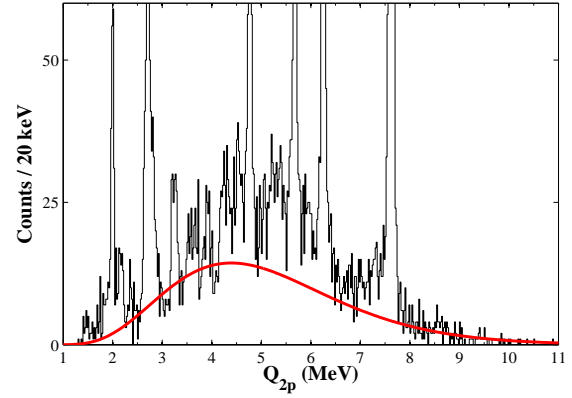}
	\caption{(Colour online) \Qtp calculated for multiplicity-two events. The upper part of the spectrum has been cut off for better visualisation of the structure, when disregarding the resolved lines. The red/grey curve is the $\chi^2$-distribution from \fref{Q1fig} folded with itself and manually scaled to the data.}%
	\label{Q2fig}%
\end{figure} 

Folding the distribution with itself gives the typical value of \Qtp. The \Qtp-spectrum can be seen in \fref{Q2fig} together with this distribution, which is again scaled to fit the data. Also here the simple model describes the data fairly well, when disregarding the resolved lines.

Finally folding the distribution with itself once more gives the typical values of \Qep. This is shown in \fref{Q3fig} together with the data. Here the statistics is very limited, but the data is consistent with the trend displayed by the simple model.

\begin{figure}[tbp]%
	\centering
	\includegraphics[width=0.95\columnwidth]{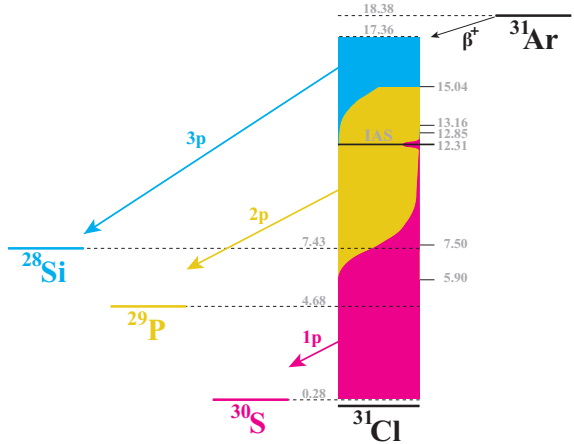}
	\caption{(Colour online) A schematic view of the decay of \Ar. The pink (dark grey), yellow (light grey) and blue (medium grey) correspond to 1p-, 2p- and 3p-decays respectively and are approximately pictured with the percentage of the proton decay as a function of excitation energy in \Cl. }%
	\label{decay}%
\end{figure}

A schematic view of the decay mode as a function of excitation energy in \Cl is shown in \fref{decay}. The approximate percentage of the decay occurring by the three decay modes is shown by colour to illustrate how they evolve as a function of excitation energy. Here it is clearly seen how the 1p-decay dominates for the low-lying levels in \Cl. When the excitation energy increases the 2p-decay takes over and finally the 3p-decay dominates for the very high excitation energies. As mentioned this is the same trend as is seen for neutron emission from nuclei close to stability.

It should be stressed that the fitted curves are only simplified models and that they are not fitted to the data, but only adjusted and scaled manually.
However, it is remarkable that with this very simple approach it is possible to relate $Q_{ip}$-distributions (for $i=1,2,3$) to each other. This relation may be exploited for other proton rich isotopes to estimate energy scales involved in $\beta$3p-decays and thereby judge if this decay mode is relevant.

\section{Summary}
\label{summary}
Using the precise energy information on the $\beta$-delayed 3p-decay of \Ar, which is obtained for the first time, we have significantly improved the determination of the high energy part of the Gamow-Teller strength distribution. The identified contribution from the 3p-decay is found to be $1.26(15)(14)$, which is of the order of 30\% of the observed strength. By comparison with a shell-model calculation we have shown that we, using our 1p-, 2p- and 3p-data, can account for the majority of the expected strength distribution.

We have shown that one-proton emission, two-proton emission and three-proton emission decay modes follow each other as dominating decay modes as the excitation energy in a nucleus is increased, as is the case for nuclei close to stability. Furthermore, we have shown that with a simplified model it is possible to give an estimate of the \Qtp- and \Qep-distributions from the \Qop-distribution, when disregarding resolved lines. 

\section{Acknowledgement}
We thank Marek Pf\"utzner for helpful discussion and input on the analysis of the $\beta$3p-decay of \Ar.

This work was supported by the European Union Seventh Framework through ENSAR (Contract No. 262010), it was partly supported by the Spanish Funding Agency under Projects No. FPA2012-32443, No. FPA2010-17142, CPAN Consolider CSD-2007-00042, and No. AIC-D-2011-0684, by the Danish Council for Independent Research Natural Sciences, by the French ANR (Contract No. ANR-06-BLAN-0320), and by R{\'e}gion Aquitaine. A.S. acknowledges support from the Jenny and Antti Wihuri Foundation.

\section*{References}
  \bibliographystyle{elsarticle-num} 
  \bibliography{ar31u}

\end{document}